# ImageR: Enhancing Bug Report Clarity by Screenshots

Xuchen Tan
York University
swithin@yorku.ca

Deenu Yadav
York University
deenuy@gmail.com

Faiz Ahmed
York University
faiz5689@yorku.ca

Maleknaz Nayebi
York University
mnayebi@yorku.ca

## ABSTRACT

In issue-tracking systems, incorporating screenshots significantly enhances the clarity of bug reports, facilitating more efficient communication and expediting issue resolution. However, determining when and what type of visual content to include remains challenging, as not all attachments effectively contribute to problem-solving; studies indicate that 22.5% of images in issue reports fail to aid in resolving the reported issues. To address this, we introduce ImageR, an AI model and tool that analyzes issue reports to assess the potential benefits of including screenshots and recommends the most pertinent types when appropriate. By proactively suggesting relevant visuals, ImageR aims to make issue reports clearer, more informative, and time-efficient.

We have curated and publicly shared a dataset comprising 6,235 Bugzilla issues, each meticulously labeled with the type of image attachment, providing a valuable resource for benchmarking and advancing research in image processing within developer communication contexts. To evaluate ImageR, we conducted empirical experiments on a subset of these reports from various Mozilla projects. The tool achieved an F1-score of 0.76 in determining when images are needed, with 75% of users finding its recommendations highly valuable. By minimizing the back-and-forth communication often needed to obtain suitable screenshots, ImageR streamlines the bug reporting process. Furthermore, it guides users in selecting the most effective visual documentation from ten established categories, potentially reducing resolution times and improving the quality of bug documentation.

ImageR is open-source, inviting further use and improvement by the community. The labeled dataset offers a rare resource for benchmarking and exploring image processing in the context of developer communication.

## 1 INTRODUCTION

Bug triaging is a critical step in software development, where developers and quality assurance (QA) teams prioritize, categorize, and assign reported software issues for resolution. Traditionally, bug reports rely heavily on textual descriptions, but these are often ambiguous, incomplete, or difficult to interpret. Recent studies indicate that incorporating images and screenshots into bug reports significantly improves their clarity and effectiveness, making it easier for developers to understand, reproduce, and fix issues efficiently [28, 29, 68, 70]. Screenshots serve as direct visual evidence of software failures, capturing error messages, UI misalignments, or unexpected behaviors that might be challenging to articulate through text alone [52]. A common interaction where a bug report is unclear through text alone and the community requests an image to assist with triaging is illustrated in Figure 1. Despite their benefits, manually analyzing images in bug reports is time-consuming and relies on human interpretation, which can introduce inconsistencies and inefficiencies in the triaging process. This has led to an increasing demand for automated tools that can interpret the content of screenshots, extract meaningful information, and support developers in prioritizing and diagnosing software issues. Recent studies show that bug reports containing visual elements are resolved 30% faster on average since screenshots can clearly convey interface issues, error states, and expected behaviors that are difficult to describe in text alone [6, 29]. However, despite proven benefits, issue-tracking systems lack robust image support. For instance, in Bugzilla, 67% of screenshots are only added after developers explicitly request them [28]. This reactive approach extends resolution times and creates communication inefficiencies. Leveraging advancements in computer vision and AI-driven image recognition, automated screenshot analysis can enhance bug triaging by identifying UI elements, recognizing error messages through optical character recognition (OCR), and even suggesting potential root causes [64, 65]. Social coding platforms, including Bugzilla [9], support image attachments but offer no automated assistance in determining or capturing such evidence. This is where Bettenburg et al.[6] conducted a study involving Mozilla developers, revealing that while screenshots significantly enhance issue clarity, there is a notable lack of tool support to optimize their use. In debugging and triage, screenshots and visuals offer additional information; however, there is limited assistance in identifying whether visual evidence would benefit a particular issue or in determining the most appropriate types of images (e.g., screenshots of error logs or workflow diagrams) [72]. In this context and motivated by the increasing use of images by developers, which was doubled between 2013 and 2018 [28] in Mozilla projects, we developed two predictive models to answer the questions below:

> When should screenshots be added to an issue report? and What types of screenshots best enhance its clarity?

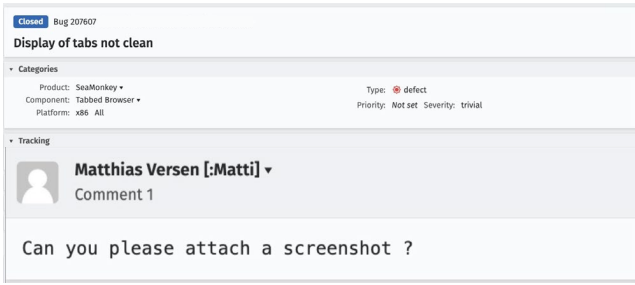

Figure 1: Example of a Bugzilla issue report where a developer requests an image attachment from the reporter.

To address this challenge, we developed an AI-driven tool that analyzes submitted bug reports and intelligently suggests the most relevant type of screenshot a developer should attach to facilitate faster and more effective triaging. Leveraging natural language processing (NLP) to assess the textual content of the report and computer vision models trained on historical bug reports, our system identifies the key visual evidence that would best support issue diagnosis. It then provides real-time guidance to reporters, recommending whether they should include a UI state screenshot, an error message capture, or a specific workflow recording based on the reported issue [64, 65]. By automating this process, we aim to reduce the back-and-forth clarifications between developers and testers, improve bug report completeness, and accelerate debugging workflows. This paper explores the role of images in bug triaging, introduces our AI-powered tool, and discusses its potential impact on software maintenance and debugging efficiency [4, 62].

### 1.1 Why an AI model is needed?

Deciding whether to include images in a bug report is typically left to the discretion of the reporter, often a developer or QA team member, who may not always recognize when visual context would improve clarity. Prior research shows that in over 73% of cases, the initial bug report lacks images, leading to follow-up comments requesting screenshots or other visual aids [30]. This reactive process not only delays issue resolution but also burdens developers with unnecessary back-and-forth communication that could have been avoided with better initial guidance.

To address this inefficiency and answer the research questions, we propose ImageR, an AI-powered model integrated into a browser extension that proactively analyzes bug report content at submission time and recommends the inclusion of relevant visual evidence. ImageR leverages supervised machine learning techniques trained on historical Bugzilla reports spanning nearly three decades (1994–2022). It uses a combination of metadata (e.g., priority, severity, and component), natural language processing (NLP) applied to textual content, and ten established screenshot categories derived from prior studies [28] to learn when and what types of screenshots are most effective.

Our approach combines classification models with sequence-to-sequence learning to generate targeted screenshot recommendations tailored to the specific content of a report. We evaluated ImageR in a user study involving 12 developers who assessed 90 real-world bug reports, measuring both the accuracy of the model's predictions and the perceived usefulness of its suggestions. Results showed that 75% of participants found the recommendations highly valuable for improving their workflow. By embedding intelligence directly into the bug reporting interface, ImageR ensures that useful visual evidence is suggested during report creation, reducing clarification delays and streamlining the overall triaging process.

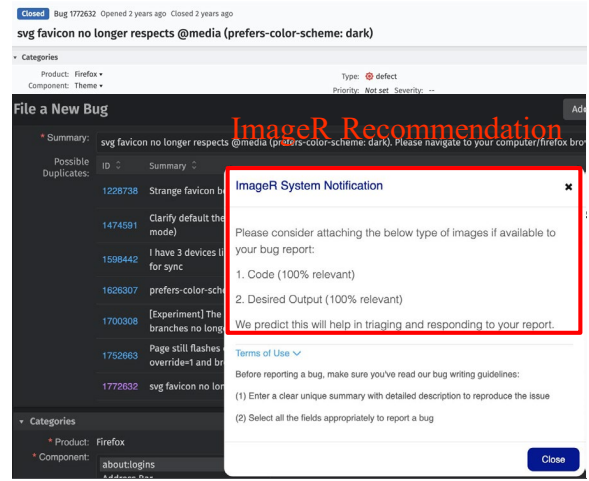

Figure 2: ImageR's recommendation to the example issue report.

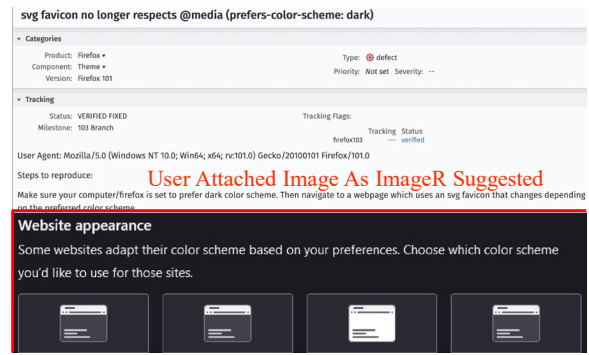

Figure 3: The enhanced issue report with ImageR's recommended image attachment of desired output.

### 1.2 Who can use this model?

ImageR is designed for software developers, quality assurance engineers, and contributors who submit bug reports on issue-tracking systems such as Bugzilla. Consider the following illustrative scenario to demonstrate how the model functions in practice.

Jane, a contributor to Mozilla projects, submits a bug report describing a layout issue in the application's UI. Her report includes a detailed textual description, but the absence of visual context leaves room for ambiguity. As a result, Bob, another developer reviewing the issue, comments and explicitly requests a screenshot to understand the problem better (see Figure 1). This common interaction introduces avoidable delays and slows down the debugging process. With ImageR, this interaction takes a different path. As Jane begins writing her report, the browser extension analyzes the text in real-time and detects that visual evidence would enhance comprehension. It prompts her with a recommendation to include a specific type of screenshot—such as a UI layout capture or a workflow illustration—tailored to the nature of the issue. ImageR ranks the usefulness of various screenshot types (e.g., error messages, interface states, configuration panels) and provides actionable guidance accordingly.

Once Jane uploads the suggested screenshot, ImageR updates the report, resulting in a more comprehensive and informative submission (see Figure 2). Enhanced reports like the one in Figure 3 proactively address common developer concerns, such as "What does the error look like?" or "Which part of the UI is affected?" By assisting users in selecting the most informative images from the outset, ImageR minimizes the need for additional clarification and follow-up requests. This not only improves communication between bug reporters and triagers but also significantly accelerates the overall resolution process. Ultimately, ImageR empowers contributors to produce more actionable reports, bridging the gap between issue identification and effective triage.

### 1.3 Paper Organization

In what follows, we first discuss related work in Section 2. We then discuss our AI model, tool's design choice, dataset, implementation, and installation instruction in Section 3, explaining how the tool works to suggest to users 'if' and 'what' kind of image they need for their issue report. We present the evaluation along with usage scenarios for the tool in Section 4. We acknowledge the limitations in Section 5 and present the conclusion in Section 6. All data and artifacts are available in the data availability statement at the end of this paper. This work builds upon the existing body of literature developed over the years [2, 3, 17, 19–21, 25–28, 31–51, 53, 56–60].

## 2 RELATED WORK

We discuss the existing AI models for bug triaging and then the use of images in developers' communication platforms.

### 2.1 AI Models for Bug Triaging

Automated bug triaging has significantly evolved over the past decade, transitioning from manual processes to sophisticated AI-driven approaches. Early methods primarily relied on classical machine learning algorithms, such as Naïve Bayes, Support Vector Machines (SVMs), and decision trees, to classify and assign bug reports based on textual descriptions. These techniques utilized features extracted from bug reports to predict the appropriate developer or team for issue resolution.

Recent advancements have seen the integration of deep learning models, including Convolutional Neural Networks (CNNs) and Long Short-Term Memory networks (LSTMs), which capture complex patterns in textual data, enhancing the accuracy of bug severity predictions and assignments. Transformer-based models, such as BERT and DeBERTa, have further improved performance by leveraging contextual embeddings to understand the nuances in bug reports [15].

Practical applications in the industry, such as the AI Q&A Copilot introduced by Razer, demonstrate the effectiveness of AI in streamlining quality assurance processes by automatically detecting and logging bugs during game testing [18]. Additionally, tools like GenaI have been developed to automate the categorization and prioritization of bug reports, reducing manual efforts and optimizing the bug triaging process [1].

### 2.2 Use of Images in the Software Engineering Context

The incorporation of images in software engineering has become increasingly prevalent, serving various purposes from enhancing developer communication to aiding in debugging and testing processes. Visual elements, such as screenshots and diagrams, improve the clarity of bug reports and documentation, facilitating better understanding among team members.

Empirical studies have highlighted the positive impact of visual information on developer productivity. For instance, research indicates that the use of images in Stack Overflow posts aids in effective communication among developers [29]. Similarly, the inclusion of visual content in GitHub issues has been shown to assist in bug diagnosis and resolution [22].

In the realm of GUI testing and debugging, vision-based tools have been developed to automate and enhance the efficiency of these processes. Tools like Sikuli utilize screenshot-based GUI scripting through template matching techniques [69], while WebDiff employs screenshot comparison for visual regression testing across different browsers [11]. VISTA leverages visual-aided search to identify UI inconsistencies in mobile applications [63], and JANUS uses video frame OCR combined with natural language processing to automate bug reproduction [13].

Table 1 provides an overview of studies and tools that highlight the use of images in various software engineering contexts.

## 3 IMAGER

ImageR is an AI model wrapped in the form of a browser extension for ease of use. ImageR assists developers in enriching their bug reports with appropriate visual evidence, thereby improving

Table 1: Overview of Image Usage and Vision-Based Tools in Software Engineering

| Image Usage in Developers' Communication | | | |
|---|---|---|---|
| **Study/Tool** | **Platform/Input Type** | **Image Type/Capability** | **Purpose/Approach** |
| Nayebi et al. [29] | Stack Overflow | Code Screenshots | Developer Communication |
| Kuramoto et al. [22] | GitHub Issues | UI Bug Screenshots | Bug Diagnosis |
| Agrawal et al. [1] | Jupyter Notebooks | Output Plots | Issue Reporting |
| Ahmed et al. [2] | Stack Overflow | Code Snapshots | Duplicate Question Detection |
| **Vision-Based Tools in GUI Testing and Debugging** | | | |
| Sikuli [69] | Screenshot | GUI Scripting | Template Matching |
| WebDiff [11] | Cross-Browser UI | Visual Regression | Screenshot Comparison |
| VISTA [63] | UI Bug Repair | GUI Images | Visual-Aided Search |
| JANUS [13] | Bug Reproduction Video | Multimodal Summarization | Video Frame OCR + NLP |
| SETU [66] | Screenshot | Joint Text-Image Similarity | Duplicate Detection |
| TANGO [12] | Video + OCR | Visual-Text Matching | Duplicate Detection |
| Wang et al. [67] | Screenshot | Multimodal Analysis | Image Utility Characterization |
| Bug Ricardo [16] | Screenshot + Logs | Heuristic Parsing + OCR | Fast Debugging |

communication efficiency and reducing resolution times. The core functionalities of ImageR include:

- **Real-Time Analysis:** As a developer drafts a bug report, ImageR analyzes the textual content in real-time to assess the need for visual aids.
- **Screenshot Recommendations:** Based on the analysis, the tool suggests specific types of screenshots—such as UI layouts, error messages, or workflow diagrams—that would best clarify the reported issue.
- **Interactive Guidance:** ImageR provides prompts and guidelines to help the reporter capture and attach the recommended visual content effectively.

The development of ImageR involved several key steps. We first started by collecting data. We curated a comprehensive dataset from historical Bugzilla reports (1994–2022), focusing on reports that included visual attachments. We then annotated the data. Each visual attachment was categorized into one of ten predefined screenshot types, based on established classifications from prior research. We then performed feature extraction. For each bug report, we extracted metadata (e.g., priority, severity, component) and performed natural language processing (NLP) on the textual descriptions to identify key features indicative of the need for visual aids. We then performed the model training. We employed supervised learning techniques, combining classification algorithms with sequence-to-sequence models, to learn patterns that suggest when and what type of screenshots would enhance issue comprehension.

## 3.1 The Imager Dataset

To support the development and evaluation of our AI model for screenshot recommendation in bug reports, we constructed the ImageR dataset—a large, curated, and labeled dataset of bug reports and image attachments from Mozilla's Bugzilla issue tracker. This dataset is designed to enable empirical research into the role of images in software issue triaging and can serve as a benchmark for future models focused on visual evidence in software maintenance.

### 3.1.1 Data Collection and Composition.

We collected 34,540 issue reports across 12 major Mozilla products from September 1, 1994, to July 19, 2022, using the Bugzilla REST API. The dataset includes two balanced subsets:

- We gathered all the bug reports with attached images. As the result, we gathered 17,270 issues with image attachments, comprising formats such as avif, jpeg, png, gif, bmp, tiff, svg+xml, and webp.
- 17,270 issues without image attachments, sampled to ensure balance for supervised learning and comparative analysis.

For each issue, we extracted metadata including *product*, *component*, *platform*, *operating system*, *severity*, *priority*, *status*, and *keywords*. We also collected the summary text and initial description fields. These underwent standard NLP preprocessing: lowercasing, stop word removal using NLTK's standard list [61], punctuation removal, and lemmatization [5]. In line with prior findings [30], we also included derived metrics such as *description length (in words)*, *number of initial comments*, and *time to first developer reply*.

For image classification, we employed ten categories established by Nayebi et al. [28]: Code, Runtime Error, Menus/Preferences, Program Input, Desired Output, Program Output, Dialog Box, Steps/Processes, CPU/GPU Performance, and Algorithm/Concept Description. We used the Amazon Mechanical Turk platform for image labeling task distribution, engaging 492 professional IT workers with more than 75% platform acceptance rates. These workers labeled 6,235 images (30% of the total 17,270 images), with each image evaluated by three workers. The research team reviewed and adjusted 695 labels (11.1% of 6,235 labeled images) through discussion and voting. Each image received a vector of ten values (0-3) indicating the frequency of assigned labels.

### 3.1.2 Image Annotation.

To classify the image attachments, we adopted the ten-category taxonomy defined by Nayebi et al. [28], including: Code, Runtime Error, Menus/Preferences, Program Input, Desired Output,

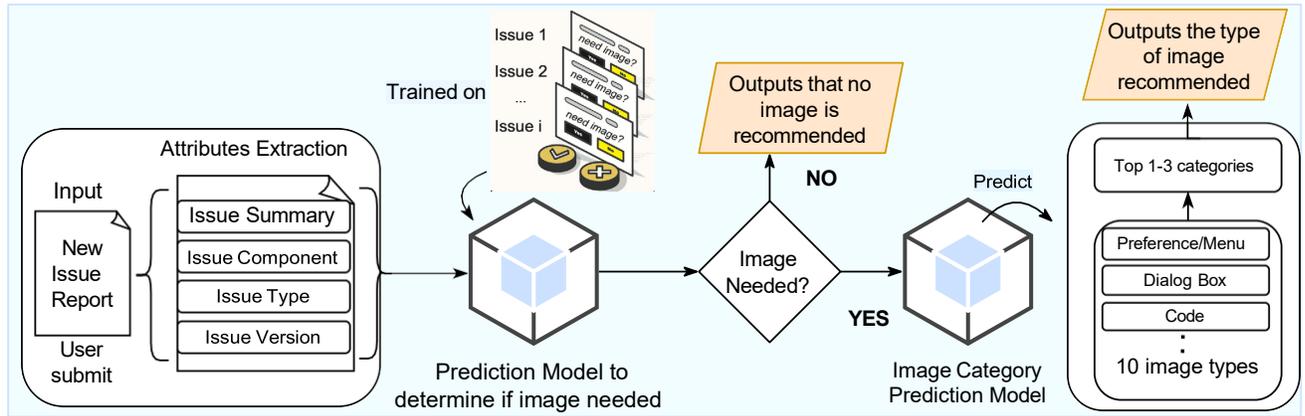

Figure 4: The technology workflow of how ImageR determines if an image attachment is needed and gives recommendations.

`Program Output`, `Steps/Processes`, `CPU/GPU Performance`, `Dialog Box`, and `Algorithm/Concept Description`.

We used Amazon Mechanical Turk to annotate a stratified sample of 6,235 images (36% of image-containing issues). Key characteristics of our labeling process:

- 492 qualified IT professionals, each with a platform approval rate of over 75%;
- Three independent annotations per image;
- Conflict resolution by majority voting, followed by manual review by our research team;
- 11.1% of labels (695 images) were refined through expert consensus to improve quality.

Each labeled image is represented as a ten-dimensional vector (0–3 scale) indicating the intensity of association with each category. This multi-label representation supports flexible classification and recommendation scenarios.

### 3.2 Architecture and Design of the AI Model

The core of `ImageR` is a two-stage AI pipeline designed to support developers during bug reporting by (1) identifying whether a screenshot is needed and (2) recommending what type of screenshot would be most helpful. This pipeline is embedded in a Chrome extension, which serves as the user-facing component. Figure 4 illustrates the workflow from user input to image recommendation.

Initially, we envisioned `ImageR` as an interactive capture tool that could dynamically access browser windows and prompt users to take and attach relevant screenshots. However, Chrome's extension architecture—designed with strong isolation and least-privilege principles—enforces strict limitations on access to browser tabs and system resources [23]. Due to these constraints, `ImageR` was implemented as a lightweight, AI-powered recommendation layer that integrates seamlessly into the Bugzilla reporting workflow without requiring privileged system access. This design ensures that developers can easily install and use the tool regardless of their familiarity with AI libraries. At the same time, the underlying models are modular and extensible, allowing more experienced developers to adapt or improve them as needed.

At the system's core, `ImageR` relies on two independently trained machine learning models that correspond to the research questions outlined in the introduction. The first model predicts whether a screenshot will likely improve a given bug report, while the second model identifies the most appropriate type(s) of image to recommend. Both models are deployed as serialized Python objects, which are loaded by the extension's backend at runtime to support real-time, context-aware recommendations.

### 3.3 Modeling Screenshot Necessity

To determine whether a screenshot should be included in a bug report, we trained a binary classifier using a combination of structured metadata and unstructured text data from our curated dataset (see Section 3.1). The model inputs include severity, priority, component, product, platform, and the preprocessed summary and description fields. Prior work suggests that image-enhanced reports are often shorter, more focused, and receive quicker responses [30], motivating the inclusion of additional features such as description length and reply count.

We trained the classifier using a Random Forest model, which was selected for its interpretability and robustness with mixed feature types. Based on our 34,540-issue dataset, each report in our training set was labeled as either image-containing or not. During prediction, the model outputs a probability score indicating the likelihood that a screenshot would benefit the issue. This score is used to drive real-time feedback to the user, prompting a recommendation when the predicted probability exceeds a learned threshold.

### 3.4 Modeling Screenshot Type Recommendation

Once a report is determined to benefit from visual support, `ImageR` triggers a second model responsible for the multi-label classification of the most appropriate screenshot categories. We adopted the ten-category taxonomy proposed by Nayebi et al. [28], covering various image types from runtime errors to workflow steps.

Each issue in our labeled training set was associated with one or more image types, as described in Section 3.1. The model receives the same feature set as the binary classifier but is trained to output a probability distribution over the ten categories. We experimented with multiple model types, including logistic regression and deep learning models; however, a linear Support Vector Machine (SVM) with one-vs-rest configuration performed best given our feature sparsity and the multi-label nature of the problem.

To generate recommendations, the model computes confidence scores for each category. The highest-scoring categories are displayed to the user via the extension interface, along with tailored text suggestions describing the ideal content of the screenshot (e.g., "Consider capturing the UI menu showing the error state").

### 3.5 Tool Integration and Runtime Behavior

The `ImageR` AI models are deployed as part of a Chrome browser extension that runs locally on the user's machine. When a developer opens the "New Bug" page on Bugzilla, the extension activates and begins analyzing text input in real-time. As the report is written, the models operate in sequence: the necessity classifier is invoked first, followed by the type predictor if visual support is deemed useful.

Despite the limitations of the browser environment, `ImageR` leverages content scripts and user-granted permissions to access the DOM of the active Bugzilla tab. It overlays non-intrusive suggestions directly in the UI, avoiding disruption to the reporting workflow. The architecture is designed to preserve privacy; no data is transmitted externally, and all predictions are performed on the client side. The models are implemented in Python 3.8, using `scikit-learn` for classification and `SpaCy` and `NLTK` for preprocessing. The training was conducted in Google Colab with GPU acceleration, and the extension backend loads pre-trained models via serialized `.pkl` files. This setup ensures end-users a lightweight, low-latency experience while maintaining reproducibility and transparency for researchers.

### 3.6 Installation

To install `ImageR`, download the entire software from our repository [71], and load it manually via Chrome's developer mode by selecting the unpacked extension from the repository. Once installed, navigate to the Bugzilla "New Bug" page [8] in Chrome. Ensure the correct version of `Python` [54] and required dependencies are installed by `requirement.txt` and run the `app.py`. Complete issue details (e.g., selecting the product, category, component, and adding descriptions), and click the extension button in your menu. `ImageR` analyzes the issue and recommends whether to include image attachments, suggesting relevant image categories with confidence scores directly in Bugzilla.

## 4 IMAGER EVALUATION

To assess the performance and practical utility of `ImageR`, we conducted both quantitative and qualitative evaluations. Our goal was to evaluate the predictive accuracy of the underlying AI models and to understand the perceived usefulness of the tool from the perspective of software developers.

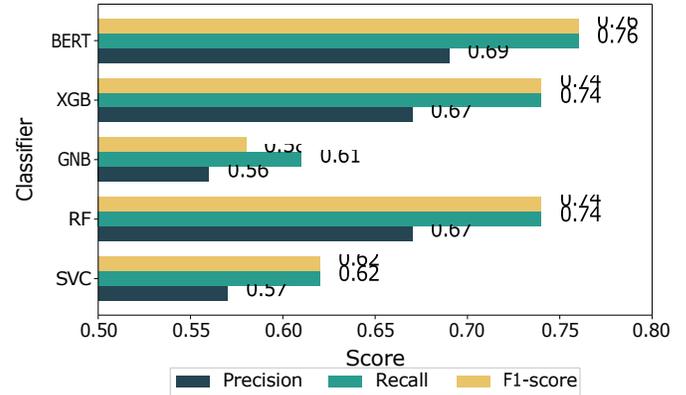

**Figure 5: Performance of Metadata-based Classification Models.**

### 4.1 Quantitative Evaluation

We began by benchmarking multiple supervised learning models for the binary classification task of predicting whether a bug report would benefit from a screenshot. Using structured metadata and preprocessed textual summaries from the `ImageR` dataset (see Section 3.1), we trained and evaluated four classifiers implemented via `scikit-learn`: Support Vector Classifier (SVC)[14], Random Forest (RF)[7], Gaussian Naive Bayes (GNB), and XGBoost (XGB) [10]. Among these, the Random Forest classifier achieved the highest F1-score of 0.74.

To improve this performance, we integrated a domain-specific language model, BugzillaBERT, into the ensemble. In uncertain cases, a conservative fallback (negative voting strategy) was applied, defaulting to "no attachment needed" to reduce false positives. This hybrid configuration raised the F1-score to 0.76, as illustrated in Figure 5.

For the multi-label classification task of recommending screenshot types, we used TF-IDF vectorization [55] for textual summaries combined with a binary relevance strategy [24] and Gaussian Naive Bayes classifiers to evaluate each of the ten image categories proposed by Nayebi et al. [42]. Each category was treated as an independent binary classification task.

This model configuration showed strong predictive performance: in 82.1% of cases, the model correctly predicted five or more relevant image categories, and in 66.9% of cases, it correctly identified more than five. On average, the model achieved 6.23 correctly predicted labels per image, demonstrating its effectiveness in handling multi-label recommendations in realistic reporting scenarios.

### 4.2 Qualitative Evaluation

To complement the algorithmic evaluation, we conducted a user study with 12 software developers. Participants were evenly split by experience: six with 1–3 years of professional development experience and six with more than three years. Each participant independently reviewed 15 bug reports (90 reports) enriched with `ImageR`'s screenshot recommendations. For each report, developers rated the recommendation quality using a five-point Likert scale.

The results indicate a high level of perceived usefulness. Across all evaluations:

- 58.8% of recommendations received "useful" (31.2%) or "very useful" (27.7%) ratings from both reviewers.
- 34.5% of recommendations received at least one "useful" rating. Only 6.7% of the cases were rated "less useful" by both evaluators, with just 3.4% showing agreement on low usefulness.

Overall, 75% of participants rated the tool as useful or very useful in supporting their triaging and debugging tasks. Only three participants considered the tool relatively useful but expressed interest in further personalization or task-specific tuning.

These findings support the effectiveness of ImageR as an intelligent aid during the issue reporting. The combination of high classification performance and positive developer feedback suggests strong potential for integration into real-world bug-tracking workflows.

## 5   LIMITATIONS

While ImageR demonstrates promising results in both predictive performance and perceived usefulness, its current version presents several limitations.

Our qualitative evaluation involved 12 developers reviewing 90 bug reports with AI-generated recommendations. While the results show positive trends, the sample size remains limited. It may not capture the full range of developer needs, project types, or issue-reporting workflows across open-source and industrial settings. Larger-scale user studies involving diverse teams, longer-term usage, and different development platforms are necessary to more fully assess the tool's effectiveness and adaptability.

The dataset used to train and evaluate ImageR consists of issue reports from Mozilla's Bugzilla instance, covering twelve primary products between 1994 and 2022. While this provides a rich longitudinal view of bug reporting behavior, it also introduces domain-specific biases. For example, practices in UI-heavy consumer applications like Firefox may not directly translate to backend systems or embedded software. Additionally, Bugzilla differs in structure and usage conventions from other platforms such as GitHub Issues or Jira. As recommended in the EASE guidelines, future versions of the dataset will include broader sampling across issue trackers to enhance generalizability and comparative reproducibility.

The current screenshot recommendation model, which combines a traditional Random Forest classifier with BugzillaBERT, achieved an F1-score of 0.76. While this result is competitive for the task, it indicates that the model still misclassifies a substantial number of cases. Similarly, although the multi-label classification of screenshot types performs well on average (6.23 correctly predicted categories per image), further improvements are needed—particularly for less frequent image types like algorithm sketches or performance dashboards. Future work could explore the use of more expressive transformer-based models fine-tuned on triage-specific language and attention-based mechanisms to better align textual tokens with visual intent.

The limitations of ImageR fall into three areas: evaluation scope, data sampling, and model performance. First, although we conducted validation with 12 developers reviewing 90 bug reports, this sample size may not fully represent the diverse needs and practices across different development contexts. Second, our data was sourced from Bugzilla, focusing on twelve main products from 1994 to 2022, which may limit the effectiveness and generalizability on other platforms or domains. Finally, although our hybrid approach using BugzillaBERT achieved an F1-score of 0.76, future work could explore alternative methods to improve recommendation accuracy.

## 6   CONCLUSION

In this study, we introduced ImageR, an AI-driven tool designed to enhance the bug triaging process by proactively recommending the inclusion of visual content in issue reports. By analyzing available issue attributes, ImageR addresses common inefficiencies in software development, where screenshots are often requested only after initial submissions.

Our empirical evaluations demonstrate that ImageR significantly improves the readability and overall effectiveness of issue reports. The tool achieved an F1-score of 0.76 in determining the necessity of image attachments, and 75% of participating users rated its recommendations as highly valuable. By minimizing back-and-forth communication to obtain suitable screenshots, ImageR streamlines the bug reporting process. Furthermore, it guides users in selecting the most effective visual documentation from ten established categories, potentially reducing resolution times and improving the quality of bug documentation. In addition to developing ImageR, we curated a comprehensive database of issue reports with image attachments. This dataset serves as a valuable resource for future research in this field, enabling further advancements in AI-assisted bug triaging and visual content recommendation systems.

By proactively suggesting relevant visual content, ImageR enhances communication efficiency among development teams, leading to more effective and timely software maintenance and debugging. The integration of such AI-driven tools holds the potential to significantly improve software quality and developer productivity in various development contexts.

## TOOL AND DATA AVAILABILITY

To ensure verifiability and transparency, we provide:

- The full dataset (anonymized) and labeling scripts.
- Preprocessing pipelines and feature extraction code,
- Environment setup (via requirements.txt) and model loading instructions.
- A replication package published on Zenodo.

The complete data and replication package of this project is available at Zenodo Repository[71]. The dataset and associated tools comply with the ethical requirements outlined in the EASE 2025 CFP, including anonymization of all personally identifiable information and full reproducibility of the pipeline.

Following the EASE 2025 track's expectations for AI model submissions, we have made the dataset, trained models, and evaluation scripts publicly available in a persistent open-access repository (see Section 3.1). All experimental configurations, including preprocessing steps, hyperparameters, and model architectures, are documented in a replication package to facilitate independent reproduction and comparative studies. The implementation adheres to FAIR principles (Findable, Accessible, Interoperable, Reusable) and includes both code-level and architectural transparency.

Although ImageR does not process personally identifiable information, it influences how developers structure issue reports and communicate within teams. Recommendations must be presented as guidance rather than mandates, preserving user agency and con- text awareness. Future iterations will explore mechanisms for user feedback loops, allowing developers to refine or disable suggestions based on project norms.